\documentclass[aps,prd,epsfig,preprint,amsmath,amssymb,superscriptaddress,showpacs,nofootinbib]{revtex4}
\usepackage{multirow}
\usepackage{graphicx,bm}
\usepackage[usenames]{color}
\usepackage{float}
\usepackage{caption}
\usepackage{subcaption}
\usepackage{slashed}
\begin{document}

\draft

\title{Search for the anomalous quartic gauge couplings through $Z\gamma$ production at $e^{-} e^{+}$ colliders}

\author{M. K\"{o}ksal\footnote{mkoksal@cumhuriyet.edu.tr}}
\affiliation{\small Department of Physics, Sivas Cumhuriyet University, 58140, Sivas, Turkiye.}

\date{\today}

\begin{abstract}

Spontaneous breaking of the $SU(2)_{L}\times U(1)_{Y}$ electroweak symmetry of the Standard Model (SM) sets the constraints on triple gauge couplings and quartic gauge couplings. Therefore, the measurement of multiboson production in $e^{-} e^{+}$ collisions allows us to directly examine the SM predictions and perform indirect investigations of new physics beyond the SM. In this paper, we concentrate the process $e^{-} e^{+} \to e^{-} Z\gamma e^{+}$ with $Z$ boson decaying to neutrinos to investigate the anomalous quartic gauge couplings using the effective Lagrangian approach at the Compact Linear Collider (CLIC). We obtain the sensitivities on the anomalous $ f_ {Ti}/\Lambda^4$ ($i=0,2,5,6,7,8,9$) couplings taking into account the systematic uncertainties of $3, 5 \%$ at $95\%$ Confidence Level for the CLIC with $\sqrt{s}=3$ TeV. Our results show that the sensitivities on some anomalous couplings without systematic errors are up to two orders of magnitude better than the current experimental limits. Considering a realistic systematic uncertainty such as $5 \%$ from possible experimental sources, the sensitivity of all anomalous quartic couplings gets worse by about $10\%$ compared to those without systematic uncertainty for the CLIC.

\end{abstract}


\vspace{5mm}

\maketitle

\section{Introduction}

The study of physics beyond the Standard Model (BSM) aims to address deficiencies in the Standard Model (SM), such as asymmetry between matter and antimatter, the strong CP problem, the origin of mass, and the nature of dark energy and dark matter. To accomplish this, BSM physics models are explored through various processes at existing and upcoming colliders. One way to search for BSM physics is by examining anomalous gauge boson couplings, including triple and quartic gauge boson couplings, which are described by the electroweak $SU(2)_{L}\times U(1)_{Y}$ gauge symmetry of the SM. These couplings can contribute to multiboson production in colliders and their precise measurements can either confirm the SM or result in the discovery of BSM physics.

Gauge boson couplings can be extended by high-dimensional operators, contributing to the anomalous triple gauge couplings and the anomalous quartic gauge couplings.
Gauge boson operators have been defined by either linear or non-linear effective Lagrangians. In the nonlinear approach, the SM gauge symmetry is conserved and realized by using the chiral Lagrangian parameterization \cite{1,2}. The anomalous triple gauge couplings and the anomalous quartic gauge couplings arise from dimension-6 operators in this approach. Dimension-6 operators usually used to analyze the anomalous quartic gauge couplings provide great convenience for comparing LEP results \cite{3}. Also, dimension-8 operators are obtained by using a linear representation of electroweak  gauge symmetry that is broken by the conventional SM Higgs mechanism. However, after the recent discovery of a new particle that is consistent with the SM Higgs boson, it becomes significant to investigate the anomalous quartic gauge couplings based on the linear effective Lagrangian.

Anomalous quartic gauge boson couplings have been studied in the literature with experimental and phenomenological studies in three different ways: triple gauge boson production, inclusive Vector Boson Scattering (VBS) with di-bosons, and exclusive di-boson production \cite{t1,t2,t4,t5,t6,t66,t8,t9,t10,t11,t12,t13,t14,t15,t16,t17,t19,a1,a2,a3,a33,a4,a5,a6,a7,a8,a9,a10,a11,a12,a13,a133,a14,a15,a16,a17,a18,a19,a20,a21,a22,a23}. The vector boson scattering processes are usually more sensitive to the anomalous quartic gauge boson couplings than triboson productions \cite{13,14}. At the lepton colliders, the gauge bosons in the final states of vector boson scattering processes are associated with very energetic leptons in the forward region with respect to the beam. On the other hand, the triboson processes are normally suppressed by $1/s$ in the s-channel propagator. For this reason, we investigate the sensitivity of the vector boson scattering process to the anomalous neutral couplings defined by dimension-8 operators at the CLIC.

In this study, we focus on high-dimensional operators that induce anomalous quartic gauge couplings but do not have any anomalous triple gauge couplings associated with them, in order to isolate the effects of the anomalous quartic gauge couplings. These operators have a minimum dimension of 8. Ref. \cite{4} shows that the existence of dimension-8 operators is required in the presence of dimension-6 operators from the Standard Model Effective Field Theory (SMEFT) space perspective. However, there are works that show the absence of contributions from dimension-6 operators \cite{5,6,7,8,9,10}. Additionally, anomalous quartic gauge couplings can lead to a greater variety of helicity combinations compared to the anomalous triple gauge couplings from dimension-6 operators \cite{11}. Furthermore, the anomalous quartic gauge couplings can be produced at tree-level, while anomalous triple gauge couplings are generated through loop diagrams \cite{12}. This means that signals of the anomalous quartic gauge couplings from dimension-8 operators may be more prominent than the anomalous triple gauge couplings from dimension-6. Therefore, we examine the process $e^- e^+ \to e^- Z \gamma e^+ $ at the CLIC to constrain the anomalous neutral quartic gauge boson couplings defined by dimension-8 effective operators. However, the anomalous neutral quartic gauge vertices within the SM framework are forbidden at tree-level and can only appear through higher-order loop diagrams. These anomalous neutral quartic gauge couplings that parametrize BSM physics can increase the $Z\gamma$ production cross section and affect the kinematic distributions of the final-state $Z$ boson and photon.

Among the proposed future particle colliders, the CLIC is one of the lepton-collider projects targeting the energies of several TeV. The CLIC has an advantage as the next-gen collider following the High-Luminosity LHC (HL-LHC) in that it offers two complementary paths for searching for new physics. The first path concentrates on examining well-known SM processes with high precision, looking for deviations from predicted behavior. Any such deviations would serve as indirect proof of new physics BSM. The second path involves direct searches for new physics phenomena, such as producing new particles. The clean environment of electron-positron collisions is ideal, both for detecting even small deviations from the properties of the SM and for detecting rare new signals. Thus, the obtained results from the CLIC may provide crucial direction for the broader field of particle physics. The CLIC is planned to operate in three different stages: $\sqrt{s}=0.38$ TeV with $L_{int}=1.0$ ab$^{-1}$, $\sqrt{s}=1.5$ TeV with $L_{int}=2.5$ ab$^{-1}$, and $\sqrt{s}=3$ TeV with $L_{int}=$5 ab$^{-1}$, respectively \cite{16}. The CLIC experiment program specifies a longitudinal polarization of $\pm80\%$ for the electron beam, while no polarized positron beam is employed \cite{pol}. The expected integrated luminosities for the CLIC with the polarized electron beams are $L_{int}=1$ ab$^{-1}$ ($P_{e^-}=80\%$), $L_{int}=4$ ab$^{-1}$ ($P_{e^-}=-80\%$), and $L_{int}=5$ ab$^{-1}$ ($P_{e^-}=0\%$) \cite{16}. However, beam polarizations increase analysis capability and reduce systematic errors. It reveals new processes by enhancing the signal or suppressing the SM processes. The use of polarized electron beams helps in increasing signal rates and minimizing unwanted background processes.

The determination of the cross-section of any process involves considering the four potential pure chiral cross-sections, taking into account the electron beam polarizations $P_{e^-}$ and positron beam polarization $P_{e^+}$ is given as follows:

\begin{eqnarray}
\label{}
\sigma\left(P_{e^-},P_{e^+}\right)=\frac{1}{4}\{\left(1+P_{e^-}\right)\left(1+P_{e^+}\right)\sigma_{RR}+\left(1-P_{e^-}\right)\left(1-P_{e^+}\right)\sigma_{LL}
\\ \nonumber
+\left(1+P_{e^-}\right)\left(1-P_{e^+}\right)\sigma_{RL}+\left(1-P_{e^-}\right)\left(1+P_{e^+}\right)\sigma_{LR}\}\,.
\end{eqnarray}

\noindent Here, $\sigma_{RR}$ denotes the cross-section for the interaction involving right-handed electron and positron beams, while $\sigma_{LL}$ represents the cross-section for the interaction involving left-handed electron and positron beams. $\sigma_{LR}$ and $\sigma_{RL}$ show the cross-sections of left-handed electron and right-handed positron beams and vice versa. The unpolarized cross-section $\sigma_0$ can be given by

\begin{equation}
\label{}
\sigma_0=\frac{1}{4}\{\sigma_{RR}+\sigma_{LL}+\sigma_{RL}+\sigma_{LR}\}\,.
\end{equation}

\section{Dimension-8 operators for the anomalous quartic couplings}

In the SM, there are no contributions from $ZZZ\gamma$, $ZZ\gamma\gamma$, and $Z\gamma\gamma\gamma$ quartic gauge boson couplings to the $Z\gamma$ production at tree level. However, the impact of new physics BSM can be studied in the $Z\gamma$ production cross-section through high-dimensional operators that define the anomalous quartic gauge boson couplings. The effective Lagrangian approach can be used to examine the effects of BSM physics in $Z\gamma$ production. Thus, the effective Lagrangian, obtained by adding dimension-8 operators to the SM Lagrangian, can be parameterized as follows:

\begin{equation}
{\cal L}_{eff}= {\cal L}_{SM} +\sum_{k=0}^1\frac{f_{S, k}}{\Lambda^4}O_{S, k} +\sum_{j=0,1,2,5,6,7,8,9}^{}\frac{f_{T, j}}{\Lambda^4}O_{T, j}+\sum_{i=0}^{7}\frac{f_{M, i}}{\Lambda^4}O_{M, i}
\end{equation}

\noindent where $\Lambda$ is the scale of new physics, $f_{S, k}$, $f_{T, j}$ and $f_{M, i}$ show coefficients of $O_{S, k}$, $O_{T, j}$ and $O_{M, i}$ that are dimension-8 operators, respectively.
These coefficients are zero in the SM prediction. According to the structure of dimension-8 operators in Eq. (3), there are three classes of genuine operators \cite{6}.

The dimension-8 operators in Eq. (3) include contributions of the Higgs boson field ($\frac{f_{S, k}}{\Lambda^4} O_{S, k}$), Gauge boson field ($\frac{f_{T, j}}{\Lambda^4} O_{T, j}$), and Higgs-Gauge
boson field ($\frac{f_{M, i}}{\Lambda^4}O_{M, i}$). The corresponding genuine operators of these contributions are given as follows:  \\

$\bullet$ Contribution of operators with covariant derivatives only (Scalar field):

\begin{eqnarray}
O_{S, 0}&=&[(D_\mu\Phi)^\dagger (D_\nu\Phi)]\times [(D^\mu\Phi)^\dagger (D^\nu\Phi)],  \\
O_{S, 1}&=&[(D_\mu\Phi)^\dagger (D^\mu\Phi)]\times [(D_\nu\Phi)^\dagger (D^\nu\Phi)],  \\
O_{S, 2}&=&[(D_\mu\Phi)^\dagger (D_\nu\Phi)]\times [(D^\nu\Phi)^\dagger (D^\mu\Phi)].
\end{eqnarray}

$\bullet$ Contribution of operators with Gauge boson field strength tensors only (Tensor field):

\begin{eqnarray}
O_{T, 0}&=&Tr[W_{\mu\nu} W^{\mu\nu}]\times Tr[W_{\alpha\beta}W^{\alpha\beta}],  \\
O_{T, 1}&=&Tr[W_{\alpha\nu} W^{\mu\beta}]\times Tr[W_{\mu\beta}W^{\alpha\nu}],  \\
O_{T, 2}&=&Tr[W_{\alpha\mu} W^{\mu\beta}]\times Tr[W_{\beta\nu}W^{\nu\alpha}],  \\
O_{T, 5}&=&Tr[W_{\mu\nu} W^{\mu\nu}]\times B_{\alpha\beta}B^{\alpha\beta},  \\
O_{T, 6}&=&Tr[W_{\alpha\nu} W^{\mu\beta}]\times B_{\mu\beta}B^{\alpha\nu},  \\
O_{T, 7}&=&Tr[W_{\alpha\mu} W^{\mu\beta}]\times B_{\beta\nu}B^{\nu\alpha},  \\
O_{T, 8}&=&B_{\mu\nu} B^{\mu\nu}B_{\alpha\beta}B^{\alpha\beta},  \\
O_{T, 9}&=&B_{\alpha\mu} B^{\mu\beta}B_{\beta\nu}B^{\nu\alpha}.
\end{eqnarray}

$\bullet$ Contribution of both operators with covariant derivatives and field strength tensors (Mixed field):

\begin{eqnarray}
O_{M, 0}&=&Tr[W_{\mu\nu} W^{\mu\nu}]\times [(D_\beta\Phi)^\dagger (D^\beta\Phi)],  \\
O_{M, 1}&=&Tr[W_{\mu\nu} W^{\nu\beta}]\times [(D_\beta\Phi)^\dagger (D^\mu\Phi)],  \\
O_{M, 2}&=&[B_{\mu\nu} B^{\mu\nu}]\times [(D_\beta\Phi)^\dagger (D^\beta\Phi)],  \\
O_{M, 3}&=&[B_{\mu\nu} B^{\nu\beta}]\times [(D_\beta\Phi)^\dagger (D^\mu\Phi)],  \\
O_{M, 4}&=&[(D_\mu\Phi)^\dagger W_{\beta\nu} (D^\mu\Phi)]\times B^{\beta\nu},  \\
O_{M, 5}&=&[(D_\mu\Phi)^\dagger W_{\beta\nu} (D^\nu\Phi)]\times B^{\beta\mu},  \\
O_{M, 7}&=&[(D_\mu\Phi)^\dagger W_{\beta\nu} W^{\beta\mu} (D^\nu\Phi)].
\end{eqnarray}

\noindent These operators in Eqs. (4)-(21) are classified regarding Higgs vs. gauge boson field content.
The subscripts $S$, $T$, $M$ correspond to scalar (or longitudinal), $T$ transversal, and $M$ mixed. These correspond
to covariant derivatives of the Higgs field for the longitudinal part and field strength tensors for the transversal
part, respectively. It is worth mentioning that the photon appears only in the transverse part because there is no direct
interaction with the scalar field. In the set of genuine operators given in Eqs. (4)-(21), $\Phi$ shows for the Higgs doublet, and the covariant derivatives of the Higgs field are given by $D_\mu\Phi=(\partial_\mu + igW^j_\mu \frac{\sigma^j}{2}
+ \frac{i}{2}g'B_\mu )\Phi$, and $\sigma^j (j=1,2,3)$ are the Pauli matrices, while $W^{\mu\nu}$ and $B^{\mu\nu}$ represents the gauge field strength tensors for $SU(2)_L$ and $U(1)_Y$.

Table I provides all the anomalous quartic gauge boson couplings modified by dimension-8 operators. As can be seen from Table I, the operators affecting anomalous neutral gauge boson couplings are $O_{T, J}$ and $O_{M, I}$. These dimension-8 operators give rise to anomalous quartic gauge boson couplings without triple gauge boson couplings. However, the best experimental limits obtained at 95$\%$ Confidence Level (C.L.) on $ f_ {T,i}/\Lambda^4$ ($i=0,2,5,6,7,8,9$) couplings
are examined through the process $pp \to  Z (\to \nu \bar{\nu}) \gamma jj \to  \nu \bar{\nu} \gamma jj$ at the LHC with $\sqrt{s}=13$ TeV with $L_{int}=139$ fb$^{-1}$ and the process $pp \to  Z (\to ll) \gamma jj \to ll \gamma jj$ at the LHC with $\sqrt{s}=13$ TeV with $L_{int}=35.9$ fb$^{-1}$\cite{den, den1}. These are given as follows

\begin{eqnarray}
-9.4<f_ {T,0}/\Lambda^4<8.4 \,(\times10^{-2} \,\text{TeV}^{-4}),
\end{eqnarray}
\begin{eqnarray}
-2.8<f_ {T,2}/\Lambda^4<2.8 \,(\times10^{-1} \,\text{TeV}^{-4}),
\end{eqnarray}
\begin{eqnarray}
-8.8<f_ {T,5}/\Lambda^4<9.9 \,(\times10^{-2} \,\text{TeV}^{-4}),
\end{eqnarray}
\begin{eqnarray}
-1.6<f_ {T,6}/\Lambda^4<1.7 \,(\,\times\text{TeV}^{-4}),
\end{eqnarray}
\begin{eqnarray}
-2.6<f_ {T,7}/\Lambda^4<2.8 \,(\times \,\text{TeV}^{-4}),
\end{eqnarray}
\begin{eqnarray}
-5.9<f_ {T,8}/\Lambda^4<5.9 \,(\times10^{-2} \,\text{TeV}^{-4}),
\end{eqnarray}
\begin{eqnarray}
-1.3<f_ {T,9}/\Lambda^4<1.3 \,(\times10^{-1} \,\text{TeV}^{-4}).
\end{eqnarray}

\begin{table}
\caption{The anomalous quartic gauge boson couplings altered with dimension-8 operators are shown with X.}
\begin{center}
\begin{tabular}{|l|c|c|c|c|c|c|c|c|c|}
\hline
& $WWWW$ & $WWZZ$ & $ZZZZ$ & $WW\gamma Z$ & $WW\gamma \gamma$ & $ZZZ\gamma$ & $ZZ\gamma \gamma$ & $Z \gamma\gamma\gamma$ & $\gamma\gamma\gamma\gamma$ \\
\hline
\cline{1-10}
$O_{S,0}$, $O_{S,1}$                     & X & X & X &   &   &   &   &   &    \\
$O_{M,0}$, $O_{M,1}$, $O_{M,7}$ & X & X & X & X & X & X & X &   &    \\
$O_{M,2}$, $O_{M,3}$, $O_{M,4}$, $O_{M,5}$ &   & X & X & X & X & X & X &   &    \\
$O_{T,0}$, $O_{T,1}$, $O_{T,2}$           & X & X & X & X & X & X & X & X & X  \\
$O_{T,5}$, $O_{T,6}$, $O_{T,7}$           &   & X & X & X & X & X & X & X & X  \\
$O_{T,8}$, $O_{T,9}$                     &   &   & X &   &   & X & X & X & X  \\
\hline
\end{tabular}
\end{center}
\end{table}

In this work, we study the sensitivities on the anomalous $ f_ {T,i}/\Lambda^4$ ($i=0,2,5,6,7,8,9$) couplings defined by dimension-8 operators related to the anomalous neutral quartic vertices through the process $e^-e^+ \to e^- Z\gamma e^+$ with $Z$ boson decaying to neutrinos at the CLIC. The anomalous quartic gauge boson couplings relevant to the process $e^{-} e^{+} \to e^{-} Z\gamma e^{+}$ are $ZZ\gamma \gamma$, $ZZZ\gamma$ and $Z \gamma\gamma\gamma$. Operators for these couplings are obtained as follows

\begin{eqnarray}
V_{ZZ\gamma \gamma,1}=F^{\mu\nu}F_{\mu\nu}Z^{\alpha}Z_{\alpha},
\end{eqnarray}
\begin{eqnarray}
V_{ZZ\gamma \gamma,2}=F^{\mu\nu}F_{\mu\alpha}Z_{\nu}Z^{\alpha},
\end{eqnarray}
\begin{eqnarray}
V_{ZZ\gamma \gamma,3}=F^{\mu\nu}F_{\mu\nu}Z^{\alpha\beta}Z_{\alpha\beta},
\end{eqnarray}
\begin{eqnarray}
V_{ZZ\gamma \gamma,4}=F^{\mu\nu}F_{\nu\alpha}Z^{\nu\alpha}Z_{\mu\nu},
\end{eqnarray}
\begin{eqnarray}
V_{ZZ\gamma \gamma,5}=F^{\mu\nu}F_{\alpha\beta}Z_{\mu\nu}Z^{\alpha\beta},
\end{eqnarray}
\begin{eqnarray}
V_{ZZ\gamma \gamma,6}=F^{\mu\beta}F_{\alpha\nu}Z^{\mu\beta}Z_{\alpha\nu},
\end{eqnarray}
\begin{eqnarray}
V_{ZZZ \gamma,1}=F^{\mu\nu}Z_{\mu\nu}Z^{\alpha}Z_{\alpha},
\end{eqnarray}
\begin{eqnarray}
V_{ZZZ \gamma,2}=F^{\mu\nu}Z_{\mu\alpha}Z_{\nu}Z^{\alpha},
\end{eqnarray}
\begin{eqnarray}
V_{ZZZ \gamma,3}=F^{\mu\nu}Z_{\mu\nu}Z^{\alpha\beta}Z_{\alpha\beta},
\end{eqnarray}
\begin{eqnarray}
V_{ZZZ \gamma,4}=F^{\mu\alpha}Z_{\mu\beta}Z^{\nu\beta}Z_{\nu\alpha},
\end{eqnarray}
\begin{eqnarray}
V_{ZZ\gamma \gamma,4}=F^{\mu\nu}F_{\mu\alpha}Z^{\alpha}Z_{\alpha},
\end{eqnarray}
\begin{eqnarray}
V_{Z\gamma \gamma \gamma,1}=F^{\mu\nu}F_{\mu\nu}F^{\alpha\beta}Z_{\alpha\beta},
\end{eqnarray}
\begin{eqnarray}
V_{Z\gamma \gamma \gamma,2}=F^{\mu\nu}F_{\nu\alpha}F^{\alpha\beta}Z_{\beta\mu}.
\end{eqnarray}
where $Z^{\mu\nu}=\partial^\mu Z^{\nu}-\partial^\nu Z^{\mu}$. The corresponding coefficients of vertices are 

\begin{eqnarray}
\alpha_{ZZ\gamma \gamma,1}=\frac{e^2 \bar{\nu}}{16\Lambda^4}\frac{1}{c_{W}s_{W}}(\frac{s_{W}}{c_{W}} f_{M0}+2\frac{c_{W}}{s_{W}} f_{M2}-f_{M4}),
\end{eqnarray}

\begin{eqnarray}
\alpha_{ZZ\gamma \gamma,2}=\frac{e^2 \bar{\nu}}{16\Lambda^4}\frac{1}{c_{W}s_{W}}(\frac{s_{W}}{2c_{W}} f_{M7}-\frac{s_{W}}{c_{W}} f_{M1}-2\frac{c_{W}}{s_{W}} f_{M3}-2f_{M5}),
\end{eqnarray}

\begin{eqnarray}
\alpha_{ZZ\gamma \gamma,3}=\frac{c_{W}^2 s_{W}^2}{2\Lambda^4}( f_{T0}+ f_{T1}-2f_{T6}+4f_{T8})+\frac{c_{W}^2 +s_{W}^2}{\Lambda^4} f_{T5},
\end{eqnarray}

\begin{eqnarray}
\alpha_{ZZ\gamma \gamma,4}=\frac{c_{W}^2 s_{W}^2}{\Lambda^4}( f_{T2}+4f_{T9})+\frac{(c_{W}^2 -s_{W}^2)^2}{2\Lambda^4} f_{T7},
\end{eqnarray}

\begin{eqnarray}
\alpha_{ZZ\gamma \gamma,5}=\frac{c_{W}^2 s_{W}^2}{\Lambda^4}( f_{T0}+ f_{T1}-2f_{T5}+4f_{T8})+\frac{(c_{W}^2 -s_{W}^2)^2}{2\Lambda^4} f_{T6},
\end{eqnarray}

\begin{eqnarray}
\alpha_{ZZ\gamma \gamma,6}=\frac{c_{W}^2 s_{W}^2}{2\Lambda^4}( f_{T2}- 2f_{T7}+4f_{T9}),
\end{eqnarray}

\begin{eqnarray}
\alpha_{ZZZ \gamma,1}=\frac{e^2 \bar{\nu}}{8\Lambda^4}\frac{1}{c_{W}s_{W}}(f_{M0}-2 f_{M2}+\frac{c_{W}^2 -s_{W}^2}{2c_{W}s_{W}}f_{M4}), 
\end{eqnarray}

\begin{eqnarray}
\alpha_{ZZZ\gamma,2}=\frac{e^2 \bar{\nu}}{8\Lambda^4}\frac{1}{c_{W}s_{W}}(2f_{M3}-f_{M1}+\frac{f_{M7}}{2}-\frac{c_{W}^2 -s_{W}^2}{2c_{W}s_{W}}f_{M5}),
\end{eqnarray}

\begin{eqnarray}
\alpha_{ZZZ\gamma,3}=\frac{c_{W}^3 s_{W}}{\Lambda^4}( f_{T0}+ f_{T1}-f_{T5}-f_{T6})+\frac{c_{W} s_{W}^3}{\Lambda^4}(f_{T5}+f_{T6}-4f_{T8}),
\end{eqnarray}

\begin{eqnarray}
\alpha_{ZZZ\gamma,4}=\frac{c_{W}^3 s_{W}}{\Lambda^4}( f_{T2}- f_{T7})+\frac{c_{W} s_{W}^3}{\Lambda^4}(f_{T7}-4f_{T9}),
\end{eqnarray}

\begin{eqnarray}
\alpha_{Z\gamma\gamma\gamma,1}=\frac{c_{W}^3 s_{W}}{\Lambda^4}( f_{T5}+ f_{T6}-4f_{T8})+\frac{c_{W} s_{W}^3}{\Lambda^4}(f_{T1}+f_{T2}-f_{T5}-f_{T6}),
\end{eqnarray}

\begin{eqnarray}
\alpha_{Z\gamma\gamma\gamma,2}=\frac{c_{W}^3 s_{W}}{\Lambda^4}( f_{T7}- 4f_{T9})+\frac{c_{W} s_{W}^3}{\Lambda^4}(f_{T2}-f_{T7}).
\end{eqnarray}

\section{Event selection and details of analysis}

In this section, we present the analysis details of the impact of dimension-8 operators on the anomalous quartic gauge boson couplings via the process $e^{-} e^{+} \to e^{-} Z\gamma e^{+}$ at the CLIC. Feynman diagrams of the anomalous neutral quartic coupling contributions to the process $e^{-} e^{+} \to e^{-} Z\gamma e^{+}$ are depicted in Fig. 1.

The final state of the process $e^{-} e^{+} \to e^{-} Z\gamma e^{+}$ consists of electron, positron, photon, and neutrinos coming from the decay of $Z$ boson. In this case, the relevant SM background processes with the same or similar final state topology
are taken into account as the processes $e^{-} e^{+} \to e^{-} Z (\to \nu \bar{\nu}) \gamma e^{+} \to e^{-} \nu \bar{\nu} \gamma e^{+}$ \, (SM).
Thus, the examined process with non-zero effective couplings is assumed as the SM contribution. The other sources for backgrounds have the same final state as signal process: $e^{-} e^{+} \to W^{-} (\to e^{-} \bar{\nu_{e}}) W^{+} (\to e^{+} \nu_{e}) \gamma \to  e^{-} \nu_{e} \bar{\nu_{e}} \gamma e^{+}$ \,($WW\gamma$), and $e^{-} e^{+} \to Z (\to e^{-}e^{+}) Z (\to \nu \bar{\nu})\gamma \to e^{-} \nu \bar{\nu} \gamma e^{+}$ \, ($ZZ \gamma$). 

Initial state radiation (ISR) is an important issue in high-energy processes, especially for lepton colliders. ISR has a significant impact on both the cross sections that include the SM and new physics. Additionally, in the case of linear colliders, the collision of incoming lepton beams results in energy loss due to Beamstrahlung. This phenomenon is expected to have a substantial influence on the distribution of effective center-of-mass energy among the colliding particles.
In this study, we evaluated the cross-section and dynamic distributions through parton-level Monte Carlo simulations using the MadGraph5\_aMC@NLO \cite{mad}. However, MadGraph5\_aMC@NLO was extended to include the effective Lagrangians of the anomalous neutral quartic couplings, which were implemented based on FeynRules \cite{aal} and the Universal FeynRules Output (UFO) framework \cite{deg}. To incorporate the effects of ISR and Beamstrahlung, we utilized MadGraph5\_aMC@NLO. Specifically, we focused on examining the cross-sections of the signal and backgrounds for the process $e^{-} e^{+} \to e^{-} Z\gamma e^{+}$ while obtaining the limit values for the anomalous couplings.

The production of events $e^{-} e^{+} \to e^{-} Z \gamma e^{+} $ with a photon in the final state, exhibiting a large transverse momentum $p_{T}^{\gamma}$, is particularly sensitive to new physics beyond the SM. Photon in the final state of the process $e^+ e^- \to e^{-} Z \gamma e^{+} $ at the CLIC have the advantage of being identifiable with high purity and efficiency. For this reason, as known, the high dimensional operators could affect $p_{T}^{\gamma}$ photon transverse momentum, especially at the region with large $p_{T}^{\gamma}$ values, which is very useful for distinguishing between signal and background events. To find distinctive properties of this process, we should examine the kinematic distributions of missing energy transverse ($\slashed{E}_T$) in the final state. Therefore, the expected number of events as a function of $\slashed{E}_T$ and $p^\gamma_T$ for the signal and relevant backgrounds of the process $e^{-} e^{+} \to e^{-} Z \gamma e^{+} $ using default cuts including the effects of ISR and Beamstrahlung at the CLIC with a center-of-mass energy of $\sqrt{s}$=3 TeV with $P_{e^-}=0\%, -80\%, 80\%$, can be seen in Figs. 2-7, respectively. Here, all couplings are taken as 1 TeV$^{-4}$. As can be seen from figures, the deviation of signal from the SM background for the anomalous quartic gauge couplings appears to be around 200 GeV for both $p_{T}^{\gamma}$ and $\slashed{E}_T$ distributions. Thus, we impose the following cuts $p_{T}^{\gamma}>200$ GeV and $\slashed{E}_T>200$ GeV as well as $|\eta^{ e^{\pm},\gamma}|<2.5$ and $p^{e^{\pm}}_T>10$ GeV for further analysis. Thus, the cuts
should be highlighted to adequately separate the signal from the SM, $Z Z\gamma$, and $WW\gamma$ backgrounds in this work. However, in order to obtain the limits on the anomalous quartic couplings, we use the kinematic selection cuts  as follows

\begin{eqnarray}
p^{e^{\pm}}_T>10 GeV ,
\end{eqnarray}

\begin{eqnarray}
|\eta^{e^{\pm},\gamma}| < 2.5
\end{eqnarray}

\begin{eqnarray}
\Delta R(e^+, e^-) > 0.4 , \Delta R(\gamma,e^\pm) > 0.4,
\end{eqnarray}

\begin{eqnarray}
\slashed{E}_T>200 GeV,
\end{eqnarray}

\begin{eqnarray}
p^\gamma_T > 200 GeV.
\end{eqnarray}

\begin{figure}[ht]
\centerline{\scalebox{0.9}{\includegraphics{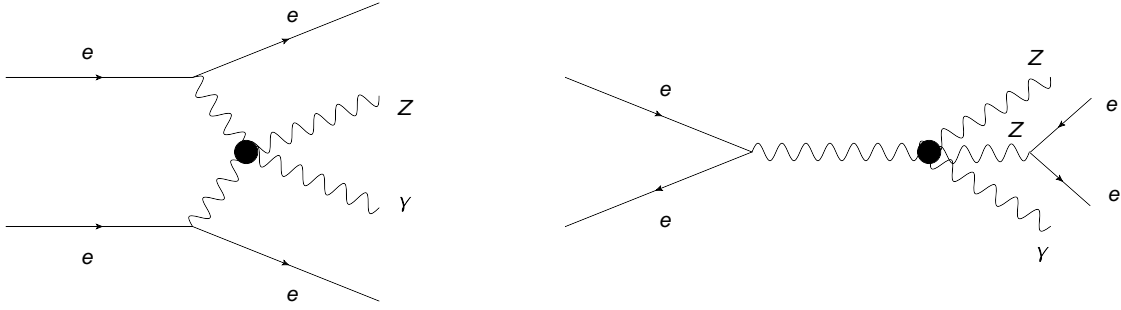}}}
\caption{Feynman diagrams of the anomalous neutral quartic couplings contribution to the process $e^{-} e^{+} \to e^{-} Z\gamma e^{+}$.}
\label{Fig.1}
\end{figure}

\begin{figure}[H]
\centerline{\scalebox{0.9}{\includegraphics{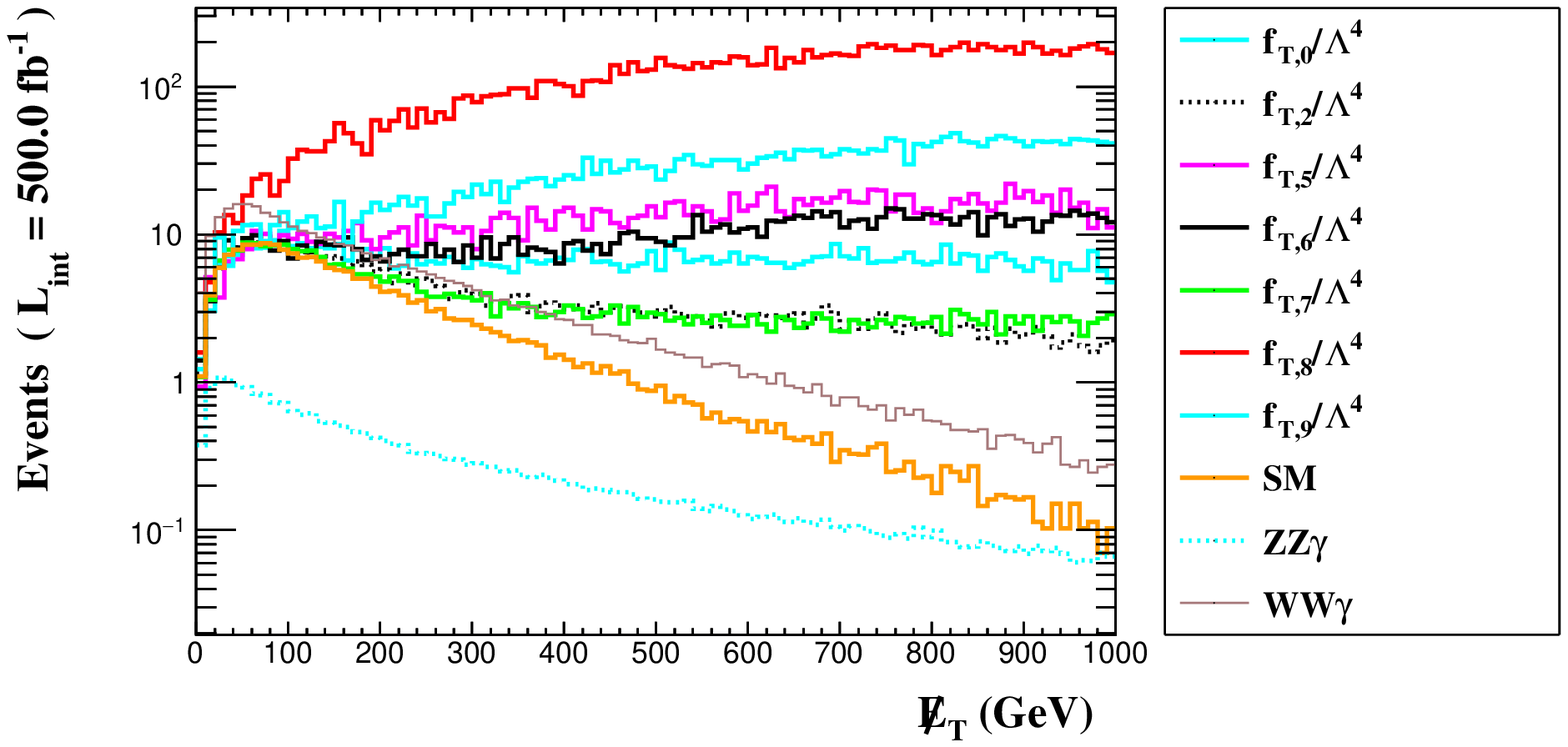}}}
\caption{ \label{fig:gamma}  The number of expected events as a function of $\slashed{E}_T$ missing energy for the $e^{-} e^{+} \to e^{-} Z\gamma e^{+} \to e^{-} \nu \bar{\nu} \gamma e^{+}$ signal and backgrounds for the unpolarized beam at the
CLIC with $\sqrt{s}=3$ TeV. The distributions are
for $f_{T,i}/\Lambda^4$ ($i=0,2,5,6,7,8,9$) and relevant backgrounds.  
Also, all couplings are taken as 1 TeV$^{-4}$.}
\end{figure}

\begin{figure}[H]
\centerline{\scalebox{0.9}{\includegraphics{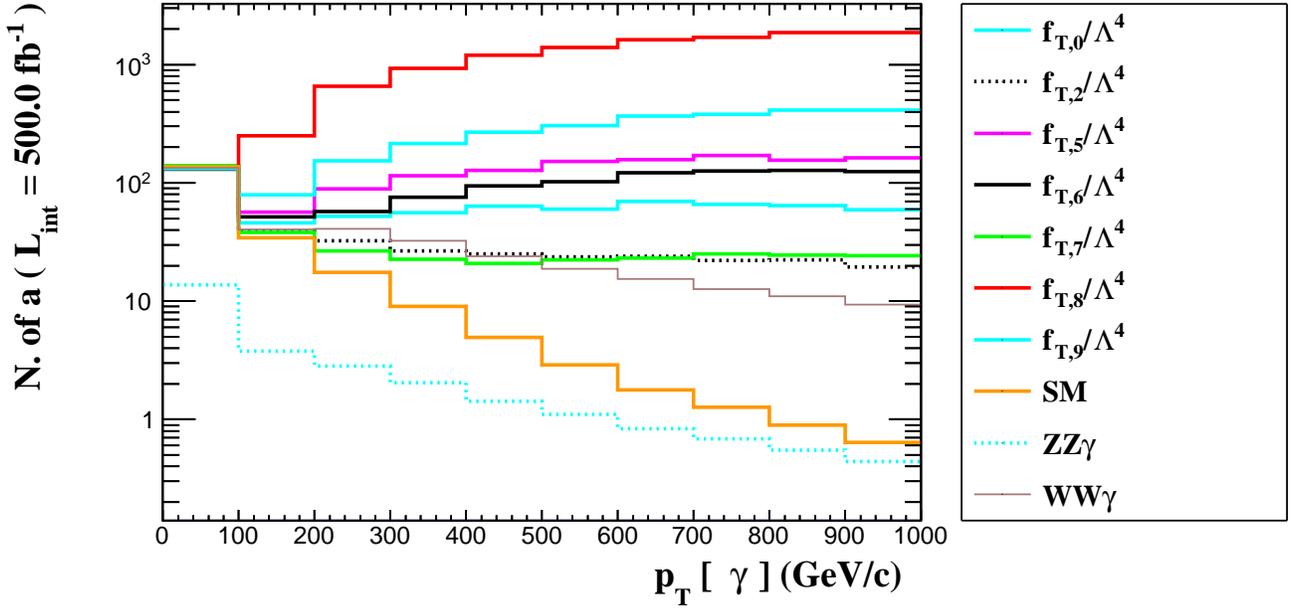}}}
\caption{ \label{fig:gamma}  The number of expected events as a function of $p^{\gamma}_T$
photon transverse momentum for the $e^{-} e^{+} \to e^{-} Z\gamma e^{+} \to e^{-} \nu \bar{\nu} \gamma e^{+}$ signal and backgrounds the unpolarized beam at the
CLIC with $\sqrt{s}=3$ TeV. The distributions are
for $f_{T,i}/\Lambda^4$ ($i=0,2,5,6,7,8,9$) and relevant backgrounds. Also, all couplings are taken as 1 TeV$^{-4}$.}
\end{figure}

\begin{figure}[H]
\centerline{\scalebox{0.9}{\includegraphics{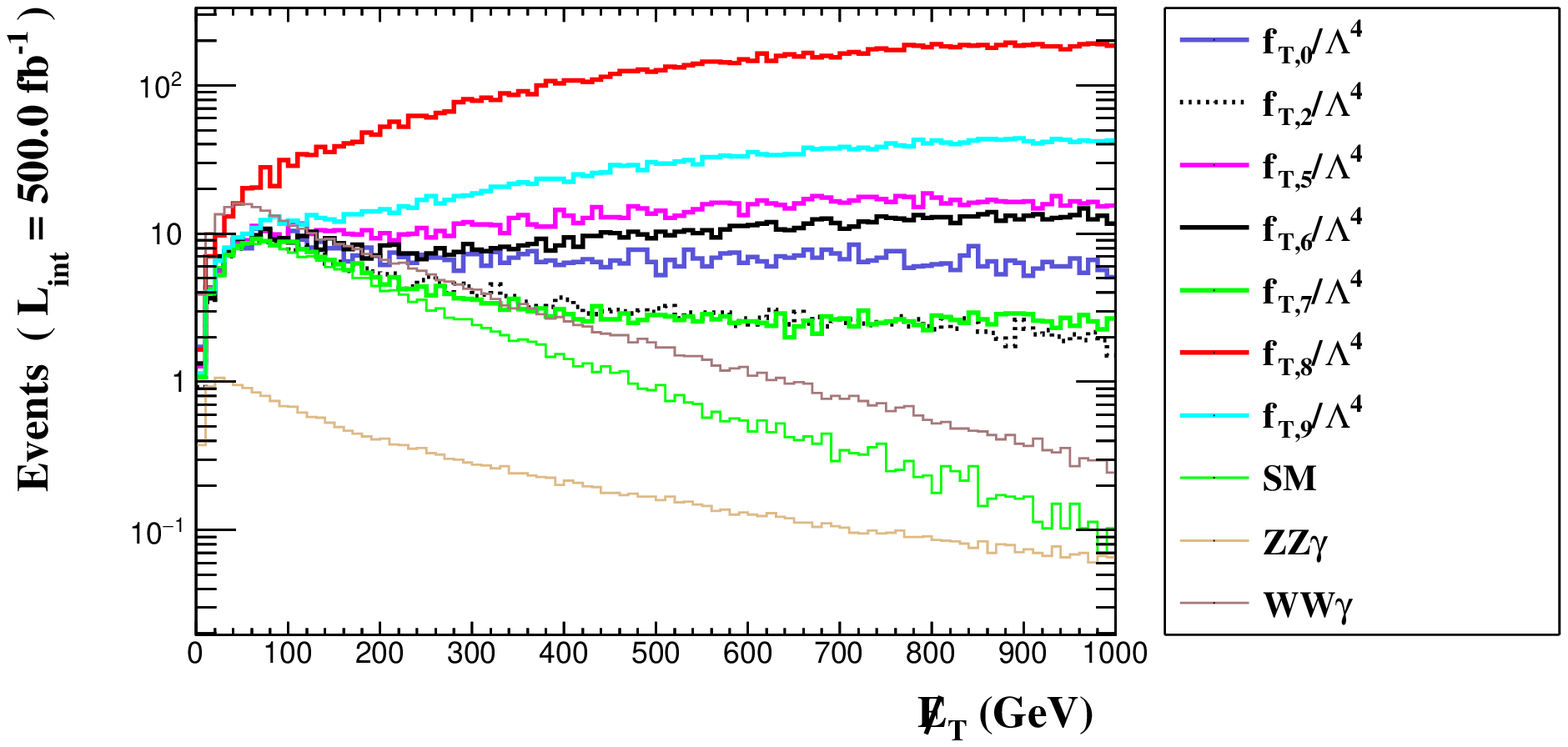}}}
\caption{ \label{fig:gamma}  Same as in Fig. 2, but for $P_{e^-}=-80\%$.}
\end{figure}

\begin{figure}[H]
\centerline{\scalebox{0.9}{\includegraphics{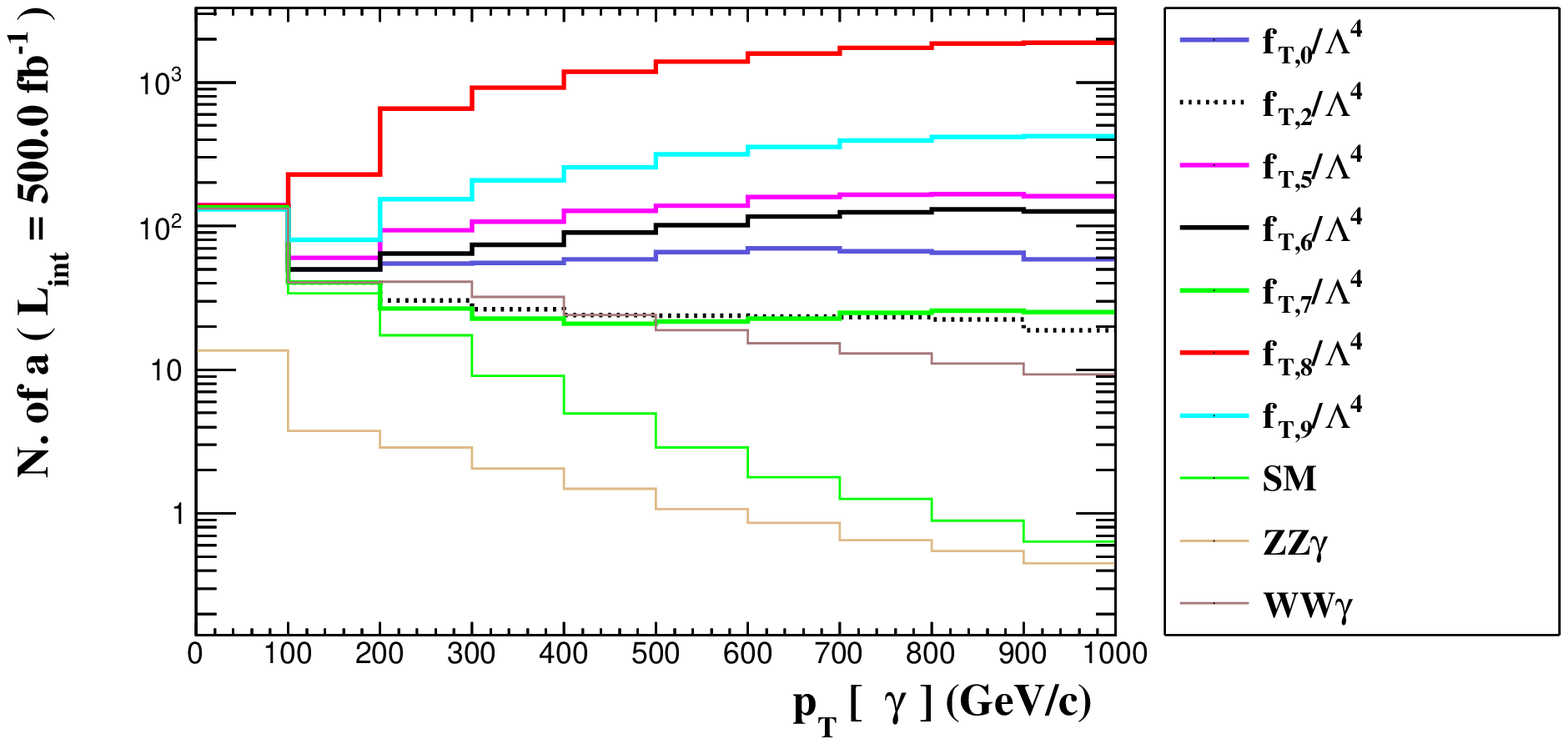}}}
\caption{ \label{fig:gamma}  Same as in Fig. 3, but for $P_{e^-}=-80\%$.}
\end{figure}

\begin{figure}[H]
\centerline{\scalebox{0.9}{\includegraphics{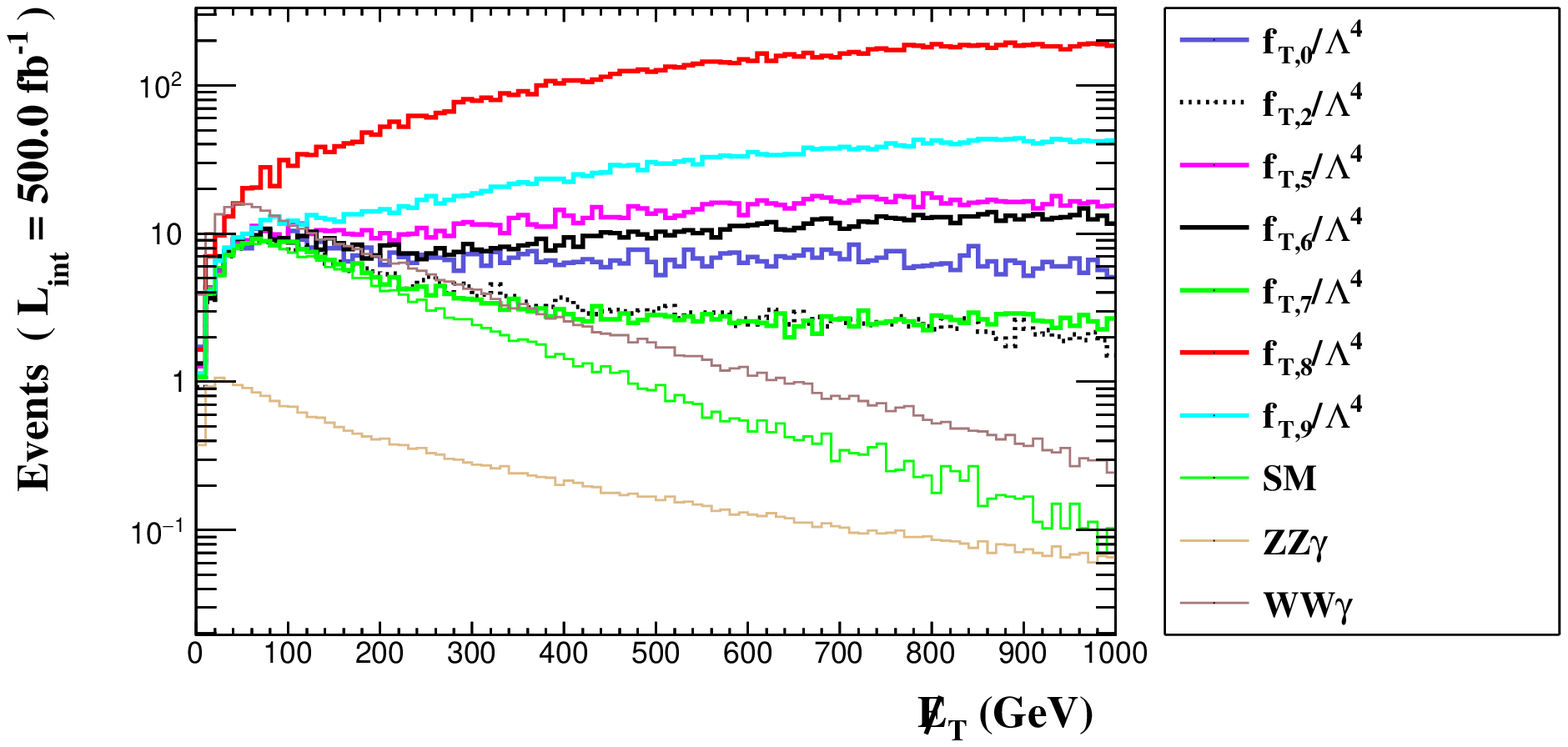}}}
\caption{ \label{fig:gamma}  Same as in Fig. 2, but for $P_{e^-}=80\%$.}
\end{figure}

\begin{figure}
\centerline{\scalebox{0.9}{\includegraphics{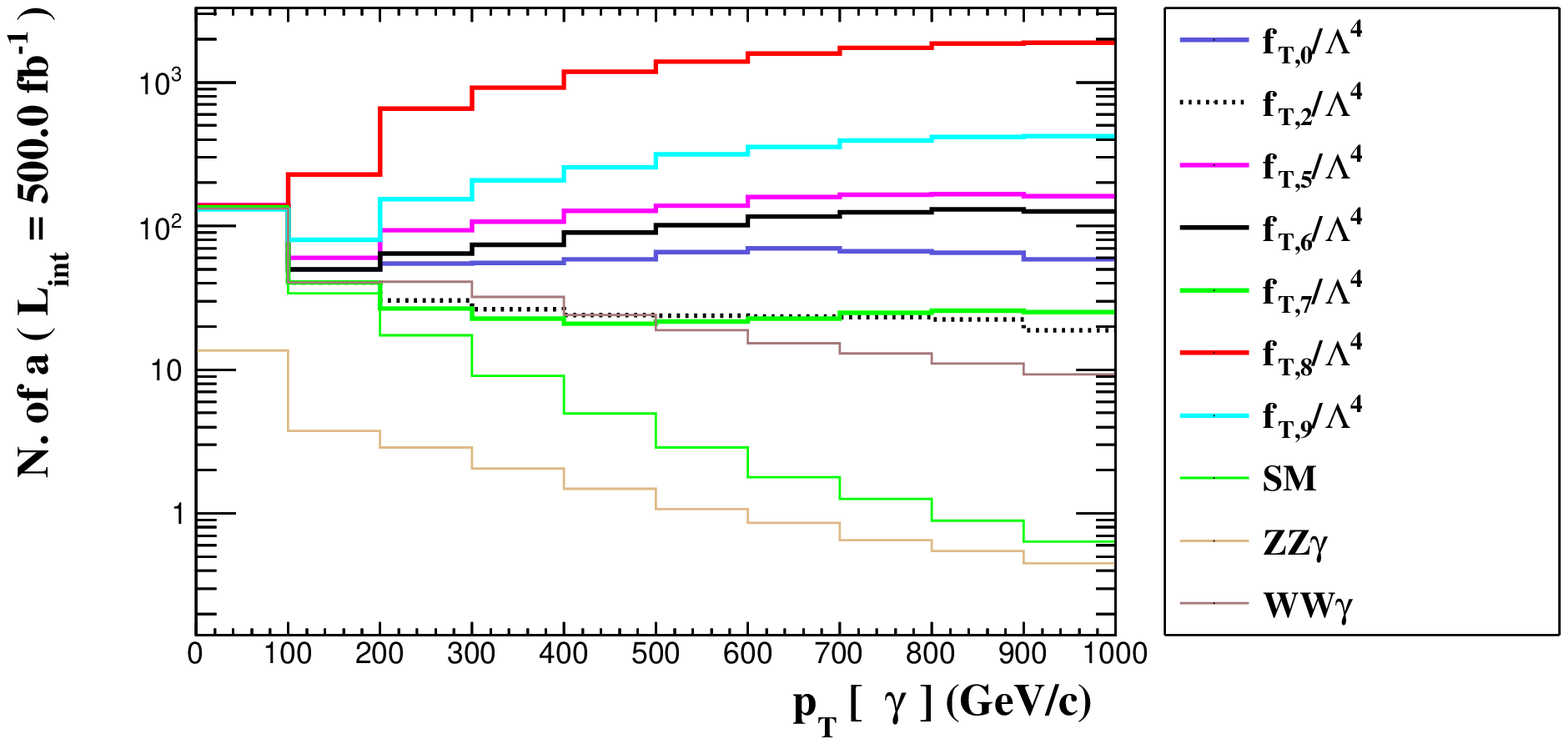}}}
\caption{ \label{fig:gamma}  Same as in Fig. 3, but for $P_{e^-}=80\%$.}
\end{figure}

\begin{figure}
\centerline{\scalebox{0.8}{\includegraphics{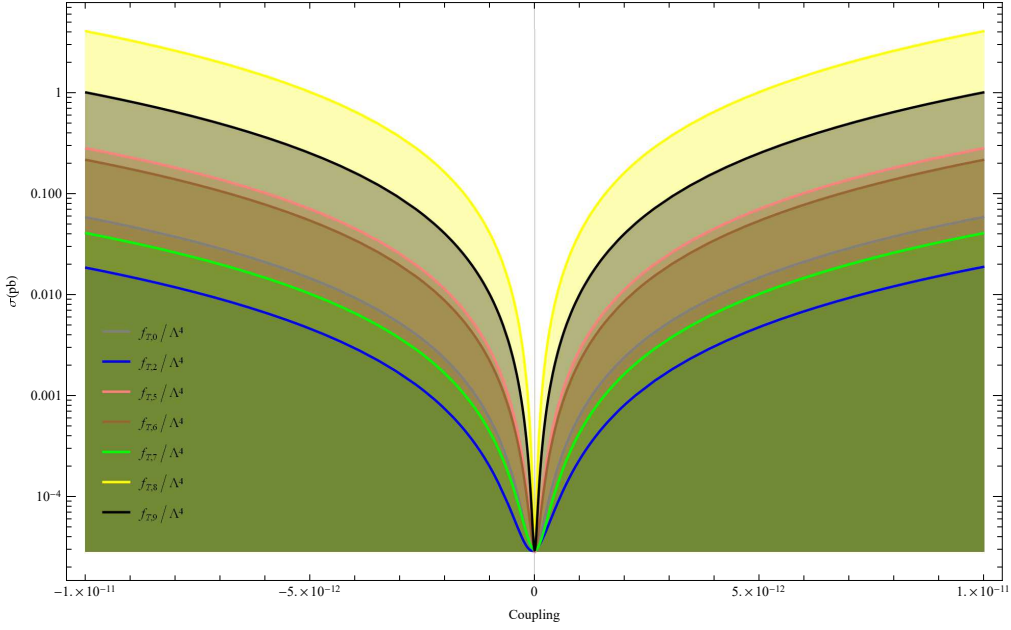}}}
\caption{ \label{fig:gamma} 
The cross sections of the $e^{-} e^{+} \to e^{-} Z\gamma e^{+}$ process as a function of the anomalous quartic gauge couplings for the CLIC.}
\end{figure}

\begin{figure}
\centerline{\scalebox{0.8}{\includegraphics{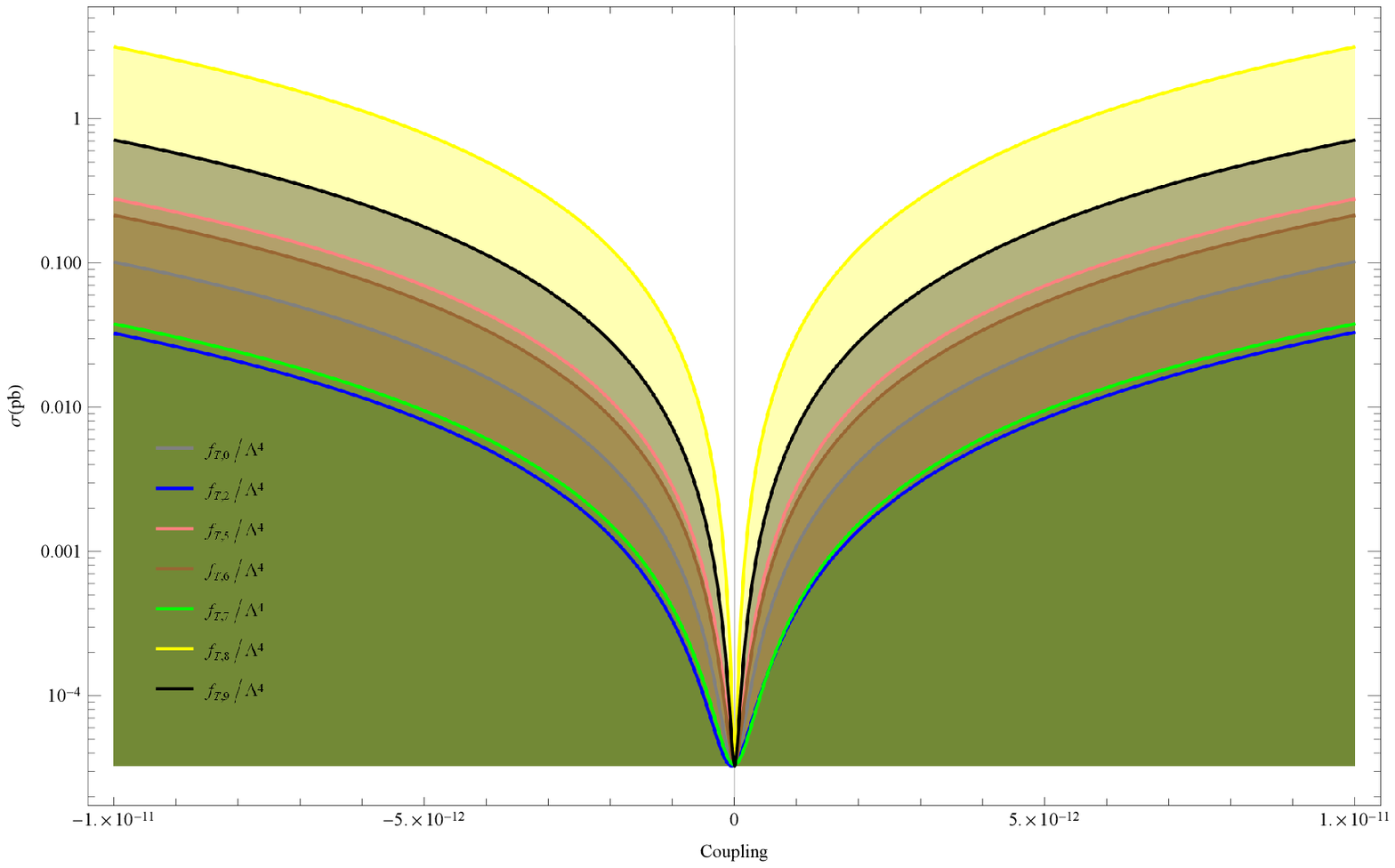}}}
\caption{ \label{fig:gamma} 
Same as in Fig. 8, but for $P_{e^-}=-80\%$.}
\end{figure}

\begin{figure}
\centerline{\scalebox{0.62}{\includegraphics{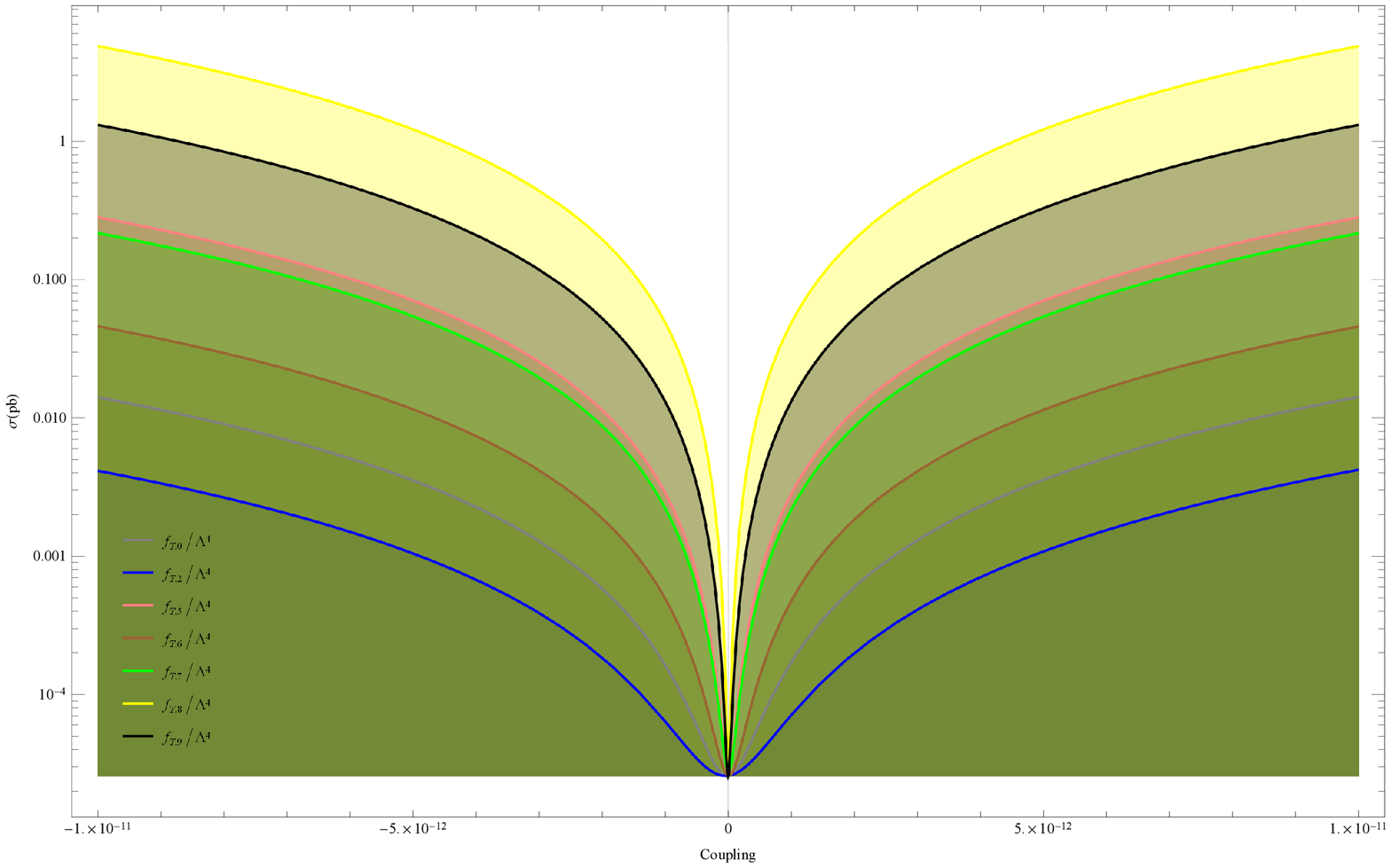}}}
\caption{ \label{fig:gamma} 
Same as in Fig. 8, but for $P_{e^-}=80\%$.}
\end{figure}

The cross sections of the $e^{-} e^{+} \to e^{-} Z\gamma e^{+}$ process as a function of the anomalous quartic gauge couplings for different polarization states at the CLIC are given in Figs. 8-10.
Also, one of the anomalous couplings is nonzero at any time, while the other couplings are fixed to zero. The total cross-sections for the $e^{-} e^{+} \to e^{-} Z\gamma e^{+}$ process increase with increasing of values of the $ f_ {T,i}/\Lambda^4$ ($i=0,2,5,6,7,8,9$) couplings. As it can be seen from Figs. 8-10, the deviation from the SM value of the anomalous
cross-section including $f_{T8}/\Lambda^4$ and $f_{T9}/\Lambda^4$ couplings is larger than that for the other couplings. 
Therefore, we expect that the limits on the anomalous $f_{T8}/\Lambda^4$ and $f_{T9}/\Lambda^4$ couplings are to be more restrictive than the obtained limits on the other couplings.

\section{SENSITIVITIES ON THE ANOMALOUS COUPLINGS}

To determine the sensitivities at a 95$\%$ C.L. for the anomalous couplings $f_{T,i}/\Lambda^4$ ($i=0,2,5,6,7,8,9$) at the CLIC, a $\chi^2$ analysis is employed, taking into account systematic errors. The $\chi^2$ function is utilized for this purpose and is defined as follows:

\begin{eqnarray}
\chi^2=\left(\frac{\sigma_{SM}-\sigma_{TOT}}{\sigma_{SM}\sqrt{\left(\delta_{st}\right)^2+\left(\delta_{sys}\right)^2}}\right)^2
\end{eqnarray}

{\raggedright where $\sigma_{SM}$ is the cross-section of relevant SM backgrounds and $\sigma_{NP}$ is given by},

\begin{eqnarray}
\sigma_{TOT}=\sigma_{SM}+\sigma_{INT}+\sigma_{BSM}.
\end{eqnarray}

{\raggedright Here,  $\sigma_{INT}$ is the interference
term between the SM and the new physics contribution and $\sigma_{BSM}$
is the contribution due to BSM physics, respectively. $\delta_{st}=\frac{1}{\sqrt {N_{SM}}}$ and $\delta_{sys}$ are the statistical error and the systematic error, respectively. The number of events of relevant SM backgrounds is given as follows:}

\begin{eqnarray}
N_{SM}=\sigma_{SM} \times L_\text{int}, 
\end{eqnarray}

{\raggedright where $L_{\text{int}}$ is the integrated luminosity.}

\begin{table} 
\caption{Expected sensitivity at $95\%$ C.L. on the anomalous quartic neutral gauge couplings via the process $e^{-} e^{+} \to e^{-} Z\gamma e^{+}$ taking into account  ISR and Beamstrahlung effects at the CLIC with $P_{e^-}=0\%, -80\%, 80\%$, $\delta_{sys}=0\%, 3\%,
5\%$ are given. Here, while obtaining the limit on any coupling is calculated, the other couplings are set to zero.}

\begin{tabular}{|c|c|c|c|c|}
\hline
$P_{e^-}$              &      & $0\%$       & $-80\%$    & $80\%$ \\
\hline
Couplings (TeV$^{-4}$) & & ${\cal L}=5$ ab$^{-1}$ & ${\cal L}=4$ ab$^{-1}$ & ${\cal L}=1$ ab$^{-1}$ \\
\hline
                      &$\delta=0\%$       &$[-11.2;7.23]\times10^{-2}$  &$[-9.74;5.69]\times10^{-2}$ &$[-2.82;2.49]\times10^{-1}$ \\
$f_{T0}/\Lambda^{4}$  &$\delta=3\%$       &$[-11.4;7.51]\times10^{-2}$  &$[-9.96;5.89]\times10^{-2}$ &$[-2.84;2.51]\times10^{-1}$ \\
                      &$\ \, \delta=5\%$ &$[-11.9;7.94]\times10^{-2}$ &$[-10.2;6.23]\times10^{-2}$ &$[-2.86;2.53]\times10^{-1}$ \\
\hline
                      &$\delta=0\%$       &$[-2.15;1.18]\times10^{-1}$  &$[-18.7;9.23]\times10^{-2}$ &$[-5.38;4.45]\times10^{-1}$ \\
$f_{T2}/\Lambda^{4}$  &$\delta=3\%$       &$[-2.20;1.23]\times10^{-1}$  &$[-19.1;9.51]\times10^{-2}$ &$[-5.42;4.47]\times10^{-1}$ \\
                      &$\ \, \delta=5\%$ &$[-2.28;1.31]\times10^{-1}$ &$[-19.6;10.1]\times10^{-2}$ &$[-5.47;4.53]\times10^{-1}$ \\
\hline
                      &$\delta=0\%$       &$[-3.98;4.17]\times10^{-2}$ &$[-4.49;4.63]\times10^{-2}$ &$[-5.73;6.10]\times10^{-2}$ \\
$f_{T5}/\Lambda^{4}$  &$\delta=3\%$       &$[-4.11;4.29]\times10^{-2}$  &$[-4.63;4.78]\times10^{-2}$ &$[-5.77;6.14]\times10^{-2}$ \\
                      &$\ \, \delta=5\%$ &$[-4.31;4.49]\times10^{-2}$ &$[-4.85;5.00]\times10^{-2}$ &$[-5.83;6.20]\times10^{-2}$ \\
\hline
                     &$\delta=0\%$       &$[-4.58;4.80]\times10^{-2}$  &$[-5.48;4.86]\times10^{-2}$ &$[-6.64;7.13]\times10^{-2}$ \\
$f_{T6}/\Lambda^{4}$ &$\delta=3\%$       &$[-4.73;4.95]\times10^{-2}$  &$[-5.63;5.01]\times10^{-2}$ &$[-6.69;7.18]\times10^{-2}$ \\
                     &$\ \, \delta=5\%$ &$[-4.97;5.12]\times10^{-2}$  &$[-5.87;5.25]\times10^{-2}$ &$[-6.76;7.25]\times10^{-2}$ \\
\hline
                      &$\delta=0\%$       &$[-1.05;1.11]\times10^{-1}$  &$[-1.28;1.16]\times10^{-1}$ &$[-1.40;1.55]\times10^{-1}$ \\
$f_{T7}/\Lambda^{4}$  &$\delta=3\%$       &$[-1.09;1.15]\times10^{-1}$  &$[-1.32;1.20]\times10^{-1}$ &$[-1.41;1.56]\times10^{-1}$ \\
                      &$\ \, \delta=5\%$ &$[-1.15;1.20]\times10^{-1}$  &$[-1.37;1.25]\times10^{-1}$ &$[-1.43;1.57]\times10^{-1}$ \\
\hline
                      &$\delta=0\%$       &$[-1.16;1.01]\times10^{-2}$  &$[-1.29;1.39]\times10^{-2}$ &  $[-1.58;1.30]\times10^{-2}$\\
$f_{T8}/\Lambda^{4}$  &$\delta=3\%$       &$[-1.20;1.05]\times10^{-2}$  &$[-1.33;1.43]\times10^{-2}$ &  $[-1.59;1.32]\times10^{-2}$\\
                      &$\ \, \delta=5\%$ &$[-1.25;1.10]\times10^{-2}$  &$[-1.39;1.49]\times10^{-2}$ &  $[-1.60;1.33]\times10^{-2}$\\
\hline
                      &$\delta=0\%$       &$[-2.41;1.96]\times10^{-2}$  &$[-2.69;2.94]\times10^{-2}$ &$[-3.23;2.35]\times10^{-2}$ \\
$f_{T9}/\Lambda^{4}$  &$\delta=3\%$       &$[-2.48;2.03]\times10^{-2}$  &$[-2.77;3.02]\times10^{-2}$ &$[-3.25;2.37]\times10^{-2}$ \\
                      &$\ \, \delta=5\%$ &$[-2.59;2.14]\times10^{-2}$  &$[-2.90;3.15]\times10^{-2}$ &$[-3.28;2.40]\times10^{-2}$ \\
\hline
\end{tabular}
\end{table}

The limits at 95$\%$ C.L. on the anomalous $ f_ {T,i}/\Lambda^4$ ($i=0,2,5,6,7,8,9$) through the process $e^{-} e^{+} \to e^{-} Z\gamma e^{+}$ at the CLIC with $\sqrt{s}$=3 TeV and various integrated luminosities are given in Table II. Our best sensitivities are presented by

\begin{eqnarray}
-9.74<f_ {T,0}/\Lambda^4<5.69 \,(\times10^{-2} \,\text{TeV}^{-4}),
\end{eqnarray}

\begin{eqnarray}
-18.7<f_ {T,2}/\Lambda^4<9.23 \,(\times10^{-2} \,\text{TeV}^{-4}),
\end{eqnarray}

\begin{eqnarray}
-3.98<f_ {T,5}/\Lambda^4<4.17 \,(\times10^{-2} \,\text{TeV}^{-4}),
\end{eqnarray}

\begin{eqnarray}
-4.58<f_ {T,6}/\Lambda^4<4.80 \,(\times10^{-2} \,\text{TeV}^{-4}),
\end{eqnarray}

\begin{eqnarray}
-1.05<f_ {T,7}/\Lambda^4<1.11 \,(\times10^{-1} \,\text{TeV}^{-4}),
\end{eqnarray}

\begin{eqnarray}
-1.16<f_ {T,8}/\Lambda^4<1.01 \,(\times10^{-2} \,\text{TeV}^{-4}),
\end{eqnarray}

\begin{eqnarray}
-2.41<f_ {T,9}/\Lambda^4<1.96 \,(\times10^{-2} \,\text{TeV}^{-4}).
\end{eqnarray}

As can be seen in Table II, the best sensitivities limits on the anomalous quartic gauge couplings at the CLIC can reach magnitudes of approximately $O(10^{1}- 10^{-2})$. Specifically, as we expected, $f_ {T,8}/\Lambda^4$ and $f_ {T,9}/\Lambda^4$ couplings have the best sensitivities among the other couplings. However, while the process $e^{-} e^{+} \to e^{-} Z\gamma e^{+}$ improves the sensitivity of $f_{T6}/\Lambda^4$ by up to a factor of $10^{2}$ compared to the LHC, sensitivities on $f_{T2}/\Lambda^4$, $f_{T7}/\Lambda^4$ and $f_{T9}/\Lambda^4$ are nearly 10 times better than the sensitivities calculated. Finally, we observed that the sensitivities obtained on $f_{T0}/\Lambda^4$, $f_{T5}/\Lambda^4$ and $f_{T8}/\Lambda^4$ are at the same order as those reported in Refs. [57-58].

We present both the case of unpolarized electron beams and the possibility of using beam polarization, which can constitute a strong advantage in searching for new physics \cite{moo}, assuming an electron beam polarization $P_{e^-}=\pm 80\%$ to correspond to the CLIC operation scenario.
In Table II, while our best sensitivities on $f_{T,0}/\Lambda^4$ and $f_{T,2}/\Lambda^4$ couplings at $P_{e^-}=-80\%$ are more restrictive than the limits obtained for the other polarizations $P_{e^-}=0\%$ and $80\%$,
$f_{T,5}/\Lambda^4$, $f_{T,6}/\Lambda^4$, $f_{T,7}/\Lambda^4$, $f_{T,8}/\Lambda^4$ and $f_{T,9}/\Lambda^4$ couplings determined at $P_{e^-}=0\%$ are better than the sensitivities derived on the anomalous quartic neutral gauge
couplings for the other polarizations. This is because the $O_{T}$ operators examined in this study have different multi-boson topologies. The limits on the anomalous quartic neutral gauge
couplings with systematic uncertainties of $3\%$ and $5\%$ are weaker than the limits obtained without systematic uncertainties. In Table II, $f_{T,8}/\Lambda^4$ coupling has the best sensitivity between the anomalous quartic neutral gauge couplings, $f_{T,8}/\Lambda^4$ without systematic uncertainty are approximately 1.03 and 1.08 times better than $3\%$ and $5\%$ systematic uncertainties limits, respectively.

\section{Conclusions}

Investigating the trilinear and quartic vector boson couplings, which are described by non-abelian gauge symmetry within the framework of the SM, can offer further validation of the model and provide valuable insights into the potential existence of new physics at a higher energy scale. This study can be parameterized using higher-order operators in the SMEFFT. Furthermore, the prospects of linear $e^{-} e^{+}$ colliders, characterized by high energy and high luminosity, present a broad opportunity for examining new physics at a higher scale. These colliders can serve as a window into exploring any potential indications of new physics, which can be parameterized using higher-order operators. Therefore, we focus on the process $e^{-} e^{+} \to e^{-} Z\gamma e^{+}$  considering ISR and Beamstrahlung effect at the CLIC to constrain the anomalous $f_{T,i}/\Lambda^4$ ($i=0,2,5,6,7,8,9$) couplings defining by dimension-8 effective operators. Nevertheless, we obtain the kinematic characteristics of signal events generated from dimension-8 effective operators and used them to develop an event selection strategy. Through this approach, we determine the limits on the anomalous $f_{T,i}/\Lambda^4$ couplings at $95\%$ C.L..  Since $O_{T,8}$ and $O_{T,9}$ give rise to anomalous quartic gauge boson couplings containing only the neutral electroweak gauge bosons among the anomalous quartic operators, $f_{T,8}/\Lambda^4$ and $f_{T,9}/\Lambda^4$ couplings provide us with the best sensitivity among the other couplings. Consequently, the CLIC offers an ideal platform to explore anomalous quartic gauge boson couplings at high energies.


\newpage

\begin{references}


\bibitem{1} W.J. Stirling, A. Werthenbach, {\it Eur. Phys. J.} {\bf C14}, 103 (2000). 
\bibitem{2} G. Belanger, F. Boudjema, {\it Phys. Lett. B} {\bf 288}, 201 (1992).
\bibitem{3} G. Belanger, F. Boudjema, Y. Kurihara, D. Perret-Gallix, A.
Semenov, Eur. Phys. J. C 13, 283 (2000).
\bibitem{t1} S. Chatrchyan et al. [CMS], Phys. Rev. D 90, no.3, 032008 (2014).
\bibitem{t2} G. Aad et al. [ATLAS], Phys. Rev. Lett. 115, no.3, 031802 (2015).
\bibitem{t4} M. Aaboud et al. [ATLAS], Eur. Phys. J. C 77, no.9, 646 (2017).
\bibitem{t5} A. M. Sirunyan et al. [CMS], Phys. Rev. D 100, no.1, 012004 (2019).
\bibitem{t6} A. Tumasyan et al. [CMS], JHEP 10, 174 (2021).
\bibitem{t66} G. Aad et al. [ATLAS], Phys. Rev. Lett. 113, no.14, 141803 (2014).
\bibitem{t8} M. Aaboud et al. [ATLAS], Phys. Rev. D 95, no.3, 032001 (2017).
\bibitem{t9} M. Aaboud et al. [ATLAS], Phys. Rev. D 96, no.1, 012007 (2017).
\bibitem{t10} A. M. Sirunyan et al. [CMS], Phys. Lett. B 774, 682-705 (2017).
\bibitem{t11} A. M. Sirunyan et al. [CMS], Phys. Rev. Lett. 120, no.8, 081801 (2018).
\bibitem{t12} A. M. Sirunyan et al. [CMS], Phys. Lett. B 798, 134985 (2019).
\bibitem{t13} A. M. Sirunyan et al. [CMS], JHEP 06, 076 (2020).
\bibitem{t14} A. M. Sirunyan et al. [CMS], Phys. Lett. B 809, 135710 (2020).
\bibitem{t15} A. M. Sirunyan et al. [CMS], Phys. Lett. B 812, 135992 (2021).
\bibitem{t16} A. M. Sirunyan et al. [CMS], Phys. Lett. B 811, 135988 (2020).
\bibitem{t17} A. Tumasyan et al. [CMS], Phys. Rev. D 104, 072001 (2021).
\bibitem{t19} M. Aaboud et al. [ATLAS], Phys. Rev. D 94, no.3, 032011 (2016).




\bibitem{a1} A. Guti\'errez-Rodr\'iguez, C. G. Honorato, J. Montano, and M. A. P\'erez, {\it Phys. Rev.} {\bf D89}, 034003 (2014).
\bibitem{a2} S. C. Inan and A. V. Kisselev, {\it JHEP} {\bf 10}, 121 (2021).
\bibitem{a3} S. C. Inan and A. V. Kisselev, {\it Eur. Phys. J.} {\bf C81}, 664 (2021).

\bibitem{a33} H. Amarkhail, S.C. İnan, A.V. Kisselev, [arXiv:2306.03653 [hep-ph]].

\bibitem{a4} Ji-Chong Yang, Yu-Chen Guo, Chong-Xing Yue, and Qing Fu, {\it Phys. Rev.} {\bf D104}, 035015 (2021).

\bibitem{a5} C. Baldenegro, S. Fichet, G. von Gersdorff and C. Royon, {\it JHEP} {\bf 06}, 142 (2017).
\bibitem{a6} E. Chapon, C. Royon and O. Kepka, {\it Phys. Rev.} {\bf D81}, 074003 (2010).

\bibitem{a7} A. Gutiérrez-Rodríguez, V. Ari, E. Gurkanli, M. Köksal, M.A. Hernández-Ruíz, J. Phys. G 49  10, 105004 (2022).


\bibitem{a8} A. Gutiérrez-Rodríguez, M.A. Hernández-Ruíz, E. Gurkanli, V. Ari, and M. Köksal, Eur.Phys. J .C {\bf D81},  3, 210 (2021).


\bibitem{a9} I. Sahin and B. Sahin, {\it Phys. Rev.} {\bf D86}, 115001 (2012).


\bibitem{a10} A. M. Sirunyan et al. [CMS], Phys. Lett. B 774, 682-705 (2017) [arXiv:1708.02812 [hep-ex]].

\bibitem{a11} O. J. P. Eboli, M. C. Gonzalez-Garcia, and S. M. Lietti, {\it Phys. Rev.} {\bf D69}, 095005 (2004).


\bibitem{a12} A. Senol, C. O. Karadeniz, K. Y. Oyulmaz, C. Helveci, and H. Denizli, Nucl.Phys. B {\bf 980},  115851 (2022).


\bibitem{a13} Y.-F. Dong, Y.-C. Mao, J.-C. Yang
[arXiv:2304.01505 [hep-ph]].

\bibitem{a133}  G. Perez, M. Sekulla, D. Zeppenfeld, Eur.Phys. J. C 78  9, 759 (2018).

\bibitem{a14} A. Senol, H. Denizli, C. Helveci, [arXiv:2303.14805 [hep-ph]].




\bibitem{a15} Shuai Zhang, Ji-Chong Yang, Yu-Chen Guo
[arXiv:2302.01274 [hep-ph]].



\bibitem{a16} Ji-Chong Yang, Xue-Ying Han, Zhi-Bin Qin, Tong Li, Yu-Chen Guo, JHEP 09 074 (2022).


\bibitem{a17} E. Gurkanli, J.Phys.G 50  1, 015002 (2023).



\bibitem{a18} V. Ari, E. Gurkanli, M. Köksal, A. Gutiérrez-Rodríguez, M.A. Hernández-Ruíz, Nucl.Phys. B 989 116133
(2023).


\bibitem{a19} E. Gürkanli, V. Ari, A. Gutiérrez-Rodríguez, M.A. Hernández-Ruíz, M. Koksal, J.Phys. G 47  9, 095006 (2020).

\bibitem{a20}  Y.-C. Guo, Y.-Y. Wang, J.-C. Yang, C.-X. Yue, Chin. Phys. C 44  12, 123105 (2020).

\bibitem{a21}  J.-C. Yang, Z.-B. Qing, X.-Y. Han, Y.-C. Guo, T. Li, JHEP 07 053 (2022).


\bibitem{a22}  Yu-Chen Guo, Y.-Y. Wang, J.-C. Yang
Nucl. Phys. B 961 115222 (2020). 

\bibitem{a23}  V. Ari, E. Gurkanli, A.A. Billur, M. Köksal
Nucl. Phys. B 957  115102 (2020).

\bibitem{13} [ATLAS Collaboration], {\it Phys. Rev. Lett.} { \bf D 450}, 115 (2015).
\bibitem{14} A. M. Sirunyan et al. [CMS], JHEP 10, 072 (2017).

\bibitem{4} C. Zhang and S. Y. Zhou, {\it Phys. Rev. Lett.} {\bf 125}, 201601 (2020).
\bibitem{5} J. Ellis and S.-F. Ge, {\it Phys. Rev. Lett.} {\bf 121}, 041801 (2018).



\bibitem{6} C. Degrande, {\it JHEP} {\bf 02}, 101 (2014).
\bibitem{7} J. Ellis, S.-F. Ge, H. J. He and R. Q. Xiao, {\it Chin. Phys. C} {\bf 44}, 063106 (2020).
\bibitem{8} J. Ellis, H. J. He and R. Q. Xiao, {\it Sci. China Phys. Mech. Astron.} {\bf 64}, 221062 (2021).
\bibitem{9} A. Senol, {\it et al.}, {\it Nucl. Phys. B} {\bf 935}, 365 (2018).
\bibitem{10} Q. Fu, J. C. Yang, C. X. Yue and Y. C. Guo, {\it Nucl. Phys. B} {\bf 972}, 115543 (2021).
\bibitem{11} G. Perez, M. Sekulla and D. Zeppenfeld, {\it Eur. Phys. J. C} {\bf 78}, 759 (2018). 
\bibitem{12} C. Arzt, M.B. Einhorn and J. Wudka, {\it Nucl. Phys. B} {\bf 433}, 41 (1995).

\bibitem{16} R. Franceschini, P. Roloff, U. Schnoor, and A. Wulzer, The Compact Linear Collider (CLIC): Physics Potential, [arXiv: 1812.07986 [hep-ex]].


\bibitem{pol} T. K. Charles, {\it et al.}, [The CLIC, CLICdp Collaborations], The Compact Linear Collider (CLIC)-2018
Summary Report, CERN-2018-005, [arXiv:1812.06018 [physics.acc-ph]].

\bibitem{den} ATLAS Collaboration, [arXiv:2208.12741 [hep-ex]].
\bibitem{den1} A. M. Sirunyan {\it et al.}, ATLAS Collaboration, {\it JHEP} {\bf 06}, 076, (2020).



\bibitem{mad} J. Alwall, M. Herquet, F. Maltoni, O. Mattelaer and T. Stelzer, {\it JHEP} {\bf 06}, 128 (2011).
\bibitem{aal} A. Alloul, N. D. Christensen, C. Degrande, C. Duhr and B. Fuks, {\it Comput. Phys. Commun.} { \bf 185} , 2250 (2014).
\bibitem{deg} C. Degrande, C. Duhr, B. Fuks, D. Grellscheid, O. Mattelaer and T. Reiter, {\it Comput. Phys. Commun.} { \bf 183}, 1201 (2012).

\bibitem{moo} G. Moortgat-Pick {\it et al.}, Phys. Rept. { \bf 460}, 131 (2008).








\end{references}
\end{document}